# The Link Between Standardization and Economic Growth:
## A Bibliometric Analysis

Jussi Heikkilä, University of Jyvaskyla, Finland & LUT University, Finland

Timo Ali-Vehmas, Independent Researcher, Finland

Julius Rissanen, Nokia Technologies, Finland

## ABSTRACT

The authors analyze the link between standardization and economic growth by systematically reviewing leading economics journals, leading economic growth researchers' articles, and economic growth-related books. They make the following observations: 1) No article has analyzed the link between standardization and economic growth in top 5 economics journals between 1996 and 2018. 2) A representative sample of the leading researchers of economic growth has allocated little attention to the link between standardization and economic growth. 3) Typically, economic growth textbooks do not contain "standards" or "standardization" in their word indices. These findings suggest that the economic growth theory has neglected the role of standardization.

## 1 INTRODUCTION

*"The lack of cooperative standardization in British industry is conspicuous in regard to locomotives. Every considerable railway has its own models, though the materials are to some extent standardized"*

Alfred Marshall (1919, p.591)

*"Standardisation and connection standards may seem purely technical details to the casual observer, but in fact they reflect the importance of achieving economies of scale."*

Nathan Rosenberg (1983, p.183)

*"Perhaps because these standards are so taken for granted, they are rarely the subject of discussion in circles beyond those in which they are formulated. They are even more rarely the subject of discussion in the public square in democratic institutions of government, or among friends. Indeed, standards are so taken for granted, so mundane, so ubiquitous, that they are extremely difficult to write about. They are usually noticed only when they fail to work."*

Lawrence Busch (2011, p.2)

Standards can be defined as rules, guidelines, or characteristics established by consensus and approved by a recognized body (see ISO/IEC, 2004). According to ISO/IEC (2004), standardization is "the process of development and application of standards" (see Choi et al., 2011). Standards are ubiquitous, and every one of us is exposed to several standards every day (Kindleberger, 1983; Busch, 2011). Consider, for instance, the measurement of time, metric systems, various safety standards, electricity standards, including plugs and sockets, data, image, video and audio compression technologies (codecs), Internet protocols, connectivity of devices via cellular networks, Wi-Fi or Bluetooth, etc. Standards have obvious public good characteristics (Kindleberger, 1983; David & Greenstein, 1990; Swann, 2000; Blind & Jungmittag, 2008) and, generally, the promotion of standards is considered beneficial, as reflected, for instance, by the increasing number of national and voluntary standards organizations and their expressed missions. For example, the International Organization for Standardization (ISO) has more than 164 national standards organizations as members that promote standardization nationally.[1]

It has been documented that societies underinvest in R&D (Jones & Williams, 2000; Lucking et al., 2018). Much less empirical evidence exists on whether societies under- or overinvest in standardization. According to Rysman and Simcoe (2009, p.1932): "the importance of SSOs has been widely discussed, yet there have been no attempts to systematically measure the effects of these institutions." Standards can be national, international, or global by their geographical dimension (Swann et al., 1996; Nadvi, 2008; Blind et al., 2018). Scale and network effects are typically greater the more international the scope of a standard is.[2] Standards have played an indispensable role, for instance, in creating and maintaining the proper functioning of the European Single Market (Pelkmans, 1987; David & Steinmueller, 1994; EC, 2018; Blind et al., 2018). Economists agree that institutions matter for economic growth (North, 1991; Mokyr, 2002; Acemoglu & Robinson, 2012). Blind and Jungmittag (2008) noted that "standards can also be interpreted as institutions. Institutional economists postulate a close relationship between institutional development and economic growth." This is also an important premise of the current article: Standards are important institutions that matter for technological progress, innovation, and, therefore, for economic growth and development.

While there exists a variety of different types of standards, we focus here on standards that are a result of open and voluntary standard development or setting and are related to technologies. We also note that the dimension of feedback processes classifies standardization organizations. Their operational mode is either one-shot standard setting or dynamic standard development (Teece, 2018).[3] Economists share the belief that innovations and technological progress are the key drivers of economic growth in the long run (Aghion & Howitt, 2009) and, presumably, technology standardization impacts the rate and direction of technological change.[4] While researchers of network economics and industrial organization economists have extensively studied standardization (e.g., Farrell & Saloner, 1985; Katz & Shapiro, 1986) and the role of patents in standard development (e.g., Lerner & Tirole, 2015), it appears that economic growth theory is almost silent about the macroeconomic impacts of standardization (e.g., Blind & Jungmittag, 2008; Swann, 2010; Baron & Schmidt, 2017). Consequently, we know much more about the microdynamics than the macrodynamics of standardization.

It is interesting that the macroeconomic impacts of technology standardization have received little attention among economists, particularly as standardization organizations have existed for more than a century.[5] Examples of standards that have had a substantial global impact include, among others, freight container standards (ISO/TC 104 Freight containers; Levinson, 2006; Bernhofen et al., 2016), Internet standards (IETF, W3C, Simcoe, 2015) and telecommunication standards (ETSI, ITU, Röller & Waverman, 2001; Teece, 2018). These standards have significantly promoted globalization and technological change. The heterogeneous nature of different standards and standardization processes makes it challenging to analyze the aggregate macroeconomic impacts of standardization. Presumably, this is a major factor explaining the dearth of research on the topic. The goal of this article is to shed more light on this research gap and the link between standardization and economic growth. We contribute to the existing literature by providing a systematic bibliometric analysis on this link and by reviewing the role of technology standardization in leading economics journals and particularly in economic growth theory.

The rest of the paper is structured as follows. Section 2 discusses the theoretical and empirical links between standardization and economic growth. Section 3 describes the theoretical framework. Methods and data are presented in Section 4, and the findings are report in Section 5. Section 6 concludes the paper.

## 2 STANDARDIZATION, TECHNOLOGICAL PROGRESS, AND ECONOMIC GROWTH

### 2.1 Theoretical Link Between Standardization and Economic Growth

The leading researchers of economic growth have analyzed a variety of factors that are associated with economic growth.[6] Economists and economic historians share the belief that technological progress is the key driver of economic growth in the long run (Solow, 1956; Mokyr, 2002; Aghion & Howitt, 2009). However, often, economists do not dig deeper into the details, or "the black box" (Rosenberg, 1983), of technological progress. Standardization has been discussed and analyzed for more than a century by economists, including Thorstein Veblen (1904) and Alfred Marshall (1919).[7] As early as 1919, Marshall discussed standardization extensively in his 1919 book, "Industry and Trade," and emphasized the important role that "multiform standardization" has had in promoting mass production in the U.S. (Langlois, 2001).[8] Yet, the paradigm of economic growth theory seems to have neglected the role of standardization in technological progress and in our increasing wellbeing.

To a large extent, the rate and direction of technological change are determined by the allocation of R&D investments, and there are several institutions that determine the incentives to invest in R&D (Arrow, 1962; Scotchmer, 2004). As mentioned, the role of institutions as determinants of economic growth has received increasing attention (North, 1990; Acemoglu & Robinson, 2012). According to North (1991), "institutions are the humanly devised constraints that structure political, economic and social interaction." By this definition, technical standards are also "institutions" (cf. Blind & Jungmittag, 2008; Featherston et al., 2016; Maze, 2017). The key message of new institutional economics is that "institutions matter" for economic growth. From the perspective of technological progress, the institutions that promote the creation and diffusion of knowledge, i.e., "efficient functioning of the knowledge infrastructure" (cf. Edquist, 1997), are the key factors. Standards are an important institution that can promote the diffusion of technologies (Blind & Jungmittag, 2008) and foster competition (Koski & Kretschmer, 2005). Standards promote interoperability, adoption of technologies, and network effects (Matutes & Regibeau, 1996; Shapiro & Varian, 1999; Swann, 2000). Thus, standards also matter for economic growth. First generation endogenous technological change models (Romer, 1990; Grossman & Helpman, 1991; Aghion & Howitt, 1992) added realism to earlier models of economic growth by treating innovation processes as endogenous. Similarly, standardization is an endogenous process that shapes technological progress.

R&D investments create positive externalities, and important progress has been made in the analysis of these knowledge spillovers (Lucking et al., 2018). Still, the quantification of externalities remains a great challenge to economists (Jones, 2016). For instance, Jones (2016) notes that there is a need for economic growth researchers to learn more about "the extent of knowledge spillovers across countries" as "each country benefits from knowledge created elsewhere in the world." Similarly, spillovers from standardization are not easily quantifiable. According to Leiponen (2008): "Opportunities to learn and accumulate social and political capabilities are thus of the essence in the creation of new standards. Firms are advised to engage in a broad cooperative approach if they wish to influence the evolution of standards." On the other hand, Blind and Mangelsdorf (2016) note: "Significant knowledge flows are apparent within standardization processes, especially from larger to the smaller German companies opposite of that seen in other types of strategic alliances. SDOs

are – as shown in Sherif (2015) for Chinese companies – interactive learning spaces. Consequently, companies' knowledge management, including their open innovation strategies, must take these opportunities into account when considering entrance into standardization."

A particular type of technology that creates large amounts of positive externalities is a "general purpose technology" (GPT). GPTs include electricity, steam, semiconductors, and the Internet (Bresnahan & Trajtenberg, 1995; Simcoe, 2015, p.16).[9] From the economic perspective, standards that promote interoperability are also particularly important (cf. Simcoe, 2015), as interoperability promotes efficiency. Standardization of technologies is closely linked to the development of general purpose technologies (Simcoe, 2015, p.26).[10] Standardized technologies, such as cellular connectivity, can be viewed as a general purpose technology (Teece, 2018) that enables downstream technological trajectories (Dosi, 1982)[11] in multiple industries (Kim et al., 2017). Standards are a heterogeneous set (Tassey, 2000; Simcoe et al., 2009; Wiegmann et al., 2017; Teece, 2018; Baron & Spulber, 2018; Baron et al., 2019), and some "upstream" standards enable a larger variety of downstream products and innovations. Swann (2000) notes: "The ultimate measure of how a standards infrastructure contributes to the economy is the sum of additional innovative products and services (and any attendant cost reductions) that grow on the back of that standards infrastructure."

Endogenous models of economic growth have incorporated competition and innovation into the analysis of economic growth, and researchers have derived stylized facts and predictions that can be explained by these dynamics (Aghion et al., 2015). However, coordination in innovation activity and how this affects the sequence of knowledge accumulation, i.e., direction of innovation, has thus far received less attention. The patent system in itself is a decentralized coordination mechanism that allocates the attention and investments of profit-maximizing agents to research projects that have the highest expected returns (Scotchmer, 2004). The public sector can also affect the rate and direction of technological change by using other "innovation policy instruments" (Takalo, 2013) or "policy toolkits" (Bloom et al., 2019), including, for example, R&D subsidies, R&D tax exemptions, and IPR systems.

Economists highlight the role of the combinatorial growth of ideas and related increasing returns (Romer, 1993; Weitzman, 1998; Jones, 2005). However, ideas-based growth cannot be only about increasing the variety of ideas because human attention and resources are limited. In a world where attention and resources are scarce, there needs to be some mechanisms that define which ideas to pursue and, also, in which sequence. The allocation of R&D investments crucially impacts the rate and direction of technological change and technological trajectories. Economic growth researchers often focus on innovation incentives but neglect incentives to create compatible and interoperable products in the value chain.

The abovementioned institutions all affect the rate and direction of technological change, but they rely mainly on competition of ideas and products in the market. Standardization, on the other hand, is based on a balance between competition and collaboration (i.e., "coopetition") where the aim is to achieve consensus (Schmidt & Werle 1998; Egyedi, 2000; Simcoe, 2015). Standardization can be understood as a coordination mechanism (Maze, 2017; Wiegmann et al., 2017) by which economic actors can collaboratively decide the technical specifications for products, etc. (Simcoe, 2015; Wiegmann et al., 2017) based on specific criteria and through either consensus or majority vote (Baron et al., 2019).[12] The focus is not on increasing the variety of ideas but, in contrast, reducing the number of ideas and amount of variety. Standardization is, in essence, variety reduction (Farrell & Saloner, 1985; Tassey, 2000). This process leads to focused technological trajectories with reduced market uncertainty (Gaynor, 2001). Sometimes, there are also competing standards (Wiegmann et al., 2017) that can co-exist and compete on the market. It is not hard to see why increasing variety of incompatible or non-interoperable product interfaces is anything but welfare increasing. In the standardization process, decision makers collectively select technical solutions on the basis of a consensus, and there is thus no need for all of the competing standards to enter the market when competition and selection of technical solutions occur within standardization process instead.

Standards can be concurrently defined as either knowledge, ideas, meta-ideas, technology, recipes, or institutions. Standardization is a process in which the technical specifications of certain aspects of technologies are codified, and thus the standard is codified knowledge. "Idea" is a very broad concept that forms the basic unit of "ideas-based growth" (Romer, 1993; Jones, 2005) and can be defined as "instructions or recipes, things that can be codified in a bitstring as a sequence of ones and zeros" (Jones, 2005). According to Romer (1993), "Perhaps the most important ideas of all are meta-ideas—ideas about how to support the production and transmission of other ideas."[13] Standards fit this definition. As already mentioned, standards are also institutions, i.e., rules of the game, as they constrain future technological development by reducing variety (cf. North, 1991). Moreover, the standardization process is a meta-idea or institution, as it is an idea about how to produce efficiently better ideas. Standards are technologies or recipes since they specify how one gets a specific output from a set of inputs (Romer, 1990). In addition, standards can be regarded as "technological trajectories" (Dosi, 1982, 1988; Cozzi, 1997; Kim et al., 2017). By reducing the variety of technologies, standardization promotes the more efficient allocation of resources as economic agents abandon some technological trajectories.

According to Grossman and Helpman (1994), "profit-seeking investments in knowledge play a critical role in the long-run growth process." Knowledge, technology, and innovations are not "manna from heaven" (cf. Audretsch, 2007). Similarly, standards are not manna from heaven. In Romer's (1990) endogenous growth model, growth "is driven by technological change that arises from intentional investment decisions made by profit-maximizing agents." Economic agents require appropriate incentives to invest in standardization activities and adopt standardized technologies. Thus, the observation that companies participate in technology standardization and adopt standardized technologies indicates that companies consider standardization to have positive expected returns.

## 2.2 Empirical Link Between Standardization and Economic Growth

The importance of standards has increased over the past decades (Lerner & Tirole, 2015), particularly in the ICT sector (Shapiro & Varian, 1999; Baron et al., 2019). According to Baron et al. (2019), standardization has long been recognized as playing an important role in technological innovation, the diffusion of new technologies, and economic growth. Therefore, it is expected that standards' impact on the rate and direction of technological progress and economic growth has similarly increased. Standards are a heterogenous set, as are standard development organizations (SDOs, Wiegmann et al., 2017; Teece, 2018; Baron & Spulber, 2018; Baron et al., 2019). Here, we briefly review whether there is any empirical link between standardization and innovation and technological change.

Schumpeterian growth theory (Aghion et al., 2015) predicts that incumbents do the most R&D. Incumbents also do the most standards development, according to empirical evidence (Larouche & Schuett, 2019). Existing empirical evidence on the impact of standards is scant (Blind & Jungmittag, 2008; Rysman & Simcoe, 2008; Baron & Schmidt, 2017). The typical challenge is the lack of data on counterfactual worlds. However, empirical evidence suggests that standardization has important impacts in multiple fields.[14]

Whereas a large share of the standardization literature has discussed the possibility that standards may lead to a lock-in to inferior technologies (David, 1985; Maze, 2017), empirical evidence suggests that SDOs perform well in selecting important technologies (Rysman & Simcoe, 2008).[15] Kim et al. (2017) find that standards are a driving force of technological convergence. Interestingly, it seems that in the economics literature, the potential negative impacts of standards seem to be sometimes more frequently cited than the benefits of standards. The anecdote of the QWERTY keyboard as an inferior de facto standard is a very popular example used to illustrate how path-dependence and standardization can lead to a lock-in to an inefficient equilibrium outcome (David, 1985).

Our understanding of the impacts of standards has expanded over time. While David (1987) identified three different purposes of the standards – (1) compatibility or interoperability, (2) minimum quality or safety, and (3) variety reduction - Swann (2010) listed eight: (1) variety reduction, (2) quality and performance, (3) measurement, (4) codified knowledge, (5) compatibility, (6) vision, (7) health and safety, and (8) environmental (see also Swann, 2015). For more extensive reviews of the impacts of standards, see Swann (2000, 2010). Network infrastructures particularly require compatibility standards (Farrell & Saloner, 1985). Telecommunication and broadband infrastructures have been found to positively affect economic growth (Röller & Waverman, 2001; Czernich et al., 2011), and telecommunication is an archetypal example of a network industry in which standards have been and continue to be subject to extensive IO analysis (Leiponen, 2008).

An important aspect of standards and standardization is geography. Standards are often classified as national, international, or global (Swann et al., 1996; Nadvi, 2008). The geographical dimension affects particularly strongly the scale effects and, in the case of compatibility and interoperability standards, the network effects of standardization. Researchers of economic growth have allocated a significant amount of their attention to country comparisons and cross-country knowledge spillovers. International trade promotes the efficient division of labor, fosters the diffusion of ideas, and causes economic growth (Frankel & Romer, 1999). Empirical evidence suggests that certain standards promote trade (Swann et al., 1996; Levinson, 2006; Bernhofen et al., 2016). Standardization increases the size of the markets and promotes economies of scale by enabling compatibility. When standards are national and differ across countries, domestic companies may face entry barriers to foreign markets due to incompatible standards. International standardization promotes, therefore, international competition, and competition puts pressure on companies to innovate. Table 1 summarizes the economic impacts of standards discussed in prior literature.

To summarize, standards are ubiquitous and their economic impact is mainly positive according to existing empirical studies. The extensive use, support, and development of standards that we observe in practice signal that standards are welfare increasing and economic growth promoting. The research question of this article is, thus, what is the role of technology standardization in leading economics journals and particularly in economic growth theory? Do leading economic growth researchers acknowledge the link between standardization and economic growth? Do books on economic growth discuss standardization? In the next section, we provide a theoretical framework for analyzing these questions.

## 3 ALLOCATION OF ATTENTION IN SCIENTIFIC RESEARCH AND EDUCATION

Why do certain research topics receive significant attention while others receive negligible attention or no attention at all? Which factors determine the set of research trajectories that we observe, and which factors determine the division of labor across scientists and the distribution of attention over the (in)finite set of research topics? How does the famous "invisible hand" generate, via a decentralized process, an equilibrium of supply and demand of scientific knowledge and enable self-interested researchers to produce scientific knowledge that meets the needs of society?[16] In this section, we present a simple framework to analyze the allocation of attention among researchers.[17] We move from general scientific progress to the specificities of economics.

### 3.1 General Framework

Science is a social institution, and researchers who produce scientific knowledge are, of course, also human beings possessing human limitations, interests, and intrinsic and extrinsic motives[18] (Goldman & Shaked, 1991; Leonard, 2002). The attention of scientists is a scarce resource and a coordination device in the production of scientific knowledge and in the development of new ideas (Klamer & van Dalen, 2001; Simcoe & Waguespack, 2011). Hence, the allocation or distribution of attention is typically skewed (Klamer & van Dalen, 2001): Some pieces of knowledge attract more cumulative attention than others. Researchers are necessarily boundedly rational decision makers, as their time and capacity to review all the existing scientific literature are limited. As a consequence,

Table 1. Economic impacts of standards, examples

| Standards by purpose | Potential positive impacts | Potential negative impacts |
|---|---|---|
| Compatibility/ Interface | Network externalities | Lock-in in old technologies if strong network externalities |
| | Avoids lock-in in old technologies | Risk of monopolization |
| | Increases variety of system products | |
| | Efficiency in supply chains | . |
| Minimum Quality/ Safety | Avoids adverse selection (Greshman´s Law) | Risk of regulatory capture; Raising rivals' costs |
| | Creates trust | Barriers to entry |
| | Reduces transaction costs | |
| Variety Reducing standards | Economics of scale | Reduces choice |
| | Focus and critical mass in emerging technologies and industries | Risk of premature selection of technologies |
| | Reduces transaction costs | |
| Information/ Measurement | Facilitates trade | |
| | Provides codified knowledge | |
| | Reduces transaction costs | |

Notes: Summary based on Swann (2000, 2010) and Blind (2013).

we understand more about those structures and behaviors of the physical world to which researchers allocate their attention.

According to Klamer and van Dalen (2001), researchers tend to cluster with likeminded researchers. There exists empirical evidence that status helps draw critical attention to a new idea (Merton, 1968; Simcoe & Waguespack, 2011). This "Matthew effect" may lead to herding and the path-dependent accumulation of scientific knowledge (Merton, 1968). There are also pecuniary rewards to received attention, and studies report a positive association between citations and salary (see Hamermesh, 2018, Table 8 for a summary).

Technological progress and standard development, as well as scientific progress and allocation of attention among researchers, are endogenous processes - that is, they are defined "within the system." The dynamics of these processes could be analyzed using rational choice theory, as would any other decision-making situation in which an optimizing decision maker with certain preferences makes choices (cf. Diamond, 1988; Brock & Durlauf, 1999). Suppose there is a finite set of researchers, an infinite set of research topics, and finite time. Researchers aim to produce scientific knowledge and compete for priority (Dasgupta & David, 1994; Strevens, 2013). Researchers' "return" would thus be conditional on being the first to create new ideas and new research results. Thus, the optimization problem faced by a researcher at a specific point in time could be characterized as a choice of distributing attention and research effort over a menu of research topics that maximizes expected utility over the researcher's career given her or his preferences, existing scientific knowledge, institutions, and beliefs over the choices of other researchers.

Irrespective of the exact form of researchers' heterogenous objective functions, preferences, and incentive systems, it is clear that researchers face tradeoffs and must make choices about how to allocate their scarce attention during the finite time that they have. The limited attention of academic researchers is necessarily focused on a specific set of topics during the finite research career.[19] Researchers must prioritize given their preferences over the set of possible research topics and also concurrently take into account and have beliefs and second-order beliefs about the choices of other competing researchers. The outcome of these correlated research topic choices and allocation of attention is a set of research paradigms and trajectories that we observe. In other words, the observed research outputs reveal researchers' preferences and beliefs about what they have considered important and worth researching.[20] These choices may have path-dependent consequences on the careers of researchers (cf. Jensen, 2013).

**Figure 1. Allocation of attention or division of cognitive labor of researchers**

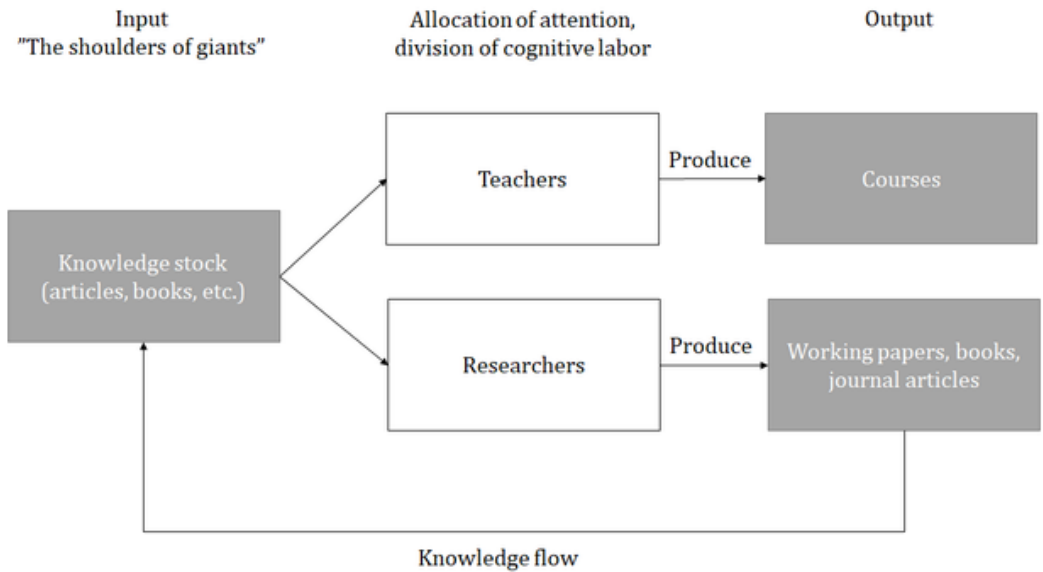

Weitzman (1998) notes that several authors have argued that invention or discovery in any sector takes place by combining ideas. Since science is cumulative by nature, there is a tendency of path-dependence and lock-in on specific topics, theories, and/or methodological choices (Akerlof, 2019). Researchers stand on the shoulders of giants and build their studies on prior existing studies and research trajectories (Kuhn, 1962; Furman & Stern, 2006). Kuhn (1962) highlighted the role of books in establishing research paradigms. He writes that "textbooks expound the body of accepted theory, illustrate many or all of its successful applications, and compare these applications with exemplary observations and experiments", and that "from textbooks each new scientific generation learns to practice its trade" (p.10). The content of textbooks and existing research articles have an impact on the framing of certain topics. Figure 1 illustrates that researchers stand on the shoulders of giants and that the existing knowledge stock defines the research topics and syllabi of courses based on prior research.

Notes: Authors' illustration.

Systematic reviews and bibliometric analyses can reveal research gaps or under-researched topics. Romer (1990) describes ideas as recipes that can be combined, and Weitzman (1998) formalized an idea-based growth model by "introducing a production function for new knowledge that depends on new recombinations of old knowledge" (see also Olsson, 2000, 2005). Research gaps can be understood as combinations of ideas that have not yet been investigated in the existing literature. Researchers have not even considered these idea combinations, or they have considered them but have instead allocated their limited attention to different, more promising topics or combinations of ideas.

### 3.2 Economics Framework

Which factors determine the set of research trajectories that we observe *in the field of economics*? How do researchers choose their research topics *in the field of economics*? Which research articles are accepted by editors to be published in leading *economics* journals? The sociology of economics research that focuses on these types of questions has a long tradition (e.g., Samuelson, 1962; Stigler & Friedman, 1975; Colander, 1989; Coupé, 2004; Hamermesh, 2018).

Empirical evidence indicates that publications and citations matter for labor market outcomes and for the salaries of economists (Coupé, 2004; Hamermesh, 2018). Therefore, economists as economic agents may focus on topics that fit well into the orthodox views of the research community and are therefore easier to publish in journals (cf. Brock & Durlauf, 1999; Akerlof, 2019; Heckman & Moktan, 2019). There can be a conflict between societal preferences and the individual preferences of economics researchers in the allocation of researchers' attention (i.e., allocative inefficiency). Even if the society might benefit from increased useful knowledge in one topic, economics researchers might still choose to focus on another topic.

Moreover, it has been documented that most novel and innovative ideas may face harsh resistance before they are accepted (Gans & Shephard, 1994), and there exists evidence of bias against novelty in science (Wang et al., 2017; Akerlof, 2019). For instance, W. Brian Arthur's seminal article (1989) was published only after multiple rejections by leading economics journals over a six-year period (Gans & Shephard, 1994). Sometimes, important new ideas are only discovered years after they are published.[21]

Hodgson and Rothman (1999), among others, have documented the dominance of a few U.S. institutions in published journal articles and journals' editorial boards. They raise the concern that such "institutional and geographical concentration of editors and authors may be unhealthy for innovative research in economics." Similarly, Coupé (2003) reports that U.S. universities and American economists dominated in the production of economics literature during 1990-2000, although the extent of their dominance did decrease over this period. Drèze and Estevan (2007, see Table 11) provide further evidence of this U.S.-oriented concentration by reporting that authors based in U.S. institutions dominate publishing in top economics journals. According to Frey and Eichenberger (1993), American economists tend to specialize in theory but neglect local institutions, whereas European economists are theoretically broad and institutionally specialized. Institutional differences between U.S. and Europe can play a role in the context of standardization research.

Presumably, researchers who focus on the theory of economic growth are no different from other researchers. Also, they build their new ideas and theories by combining ideas generated by past economic growth theory researchers. In the context of economic growth theory (and economics more generally), recipients of the Nobel Memorial Prize in Economics, Robert Solow and Paul Romer, among others, could be described as "giants" upon whose shoulders current and future growth theorists stand. The high number of forward citations received by their research articles is a clear indication of this (see Table 4).

As the curricula of economics courses themselves comprise an institution that directs the attention of economics students, it is important to analyze their content. A conscious or unconscious choice to leave some important topics, such as the economic impact of standards, out of economics curricula is likely to have certain outcomes. Students probably less often write their theses about those topics that are not discussed in their courses. The process is endogenous and incremental: Economics students become experts mainly in the fields in which they receive their education.

To our knowledge, there are no bibliometric analyses on how economists have allocated their attention to standardization and particularly on the role of standardization in the context of economic growth. Narayanan and Chen (2012) summarized the primary research streams and key arguments of technology standards research but did not focus on the perspectives of economics literature. They concluded that "the greatest opportunity lies in integrative works that will take us one step closer to a comprehensive view of technology standards." Choi et al. (2011) documented, using a bibliometric analysis, that standardization and innovation research has continuously increased over time: In 1995, there were 13 published articles focusing on these topics; in 2008, there were 68 articles on these topics (altogether, 528 articles in Web of Science). Choi et al. (2011) identified six subject-group domains of management, economics, environment, chemistry, computer science, and telecommunications, and suggested that future studies could more deeply analyze the details of these subject-group domains. This article focuses on the economics domain and seeks to offer one integrative view by analyzing the link between standardization and economic growth.

## 4 METHODOLOGY AND DATA

In order to analyze the attention allocated to the link between standardization and economic growth in the economics literature, we conduct a bibliometric analysis that uses a variety of different search techniques and methods. The analysis is purely descriptive and comprises three sections: (1) Standardization in leading economics journals, (2) Standardization in the peer-reviewed articles of leading researchers of economic growth, and (3) Standardization in a set of economic growth-related books. Scientific articles and books comprise the core of the knowledge base that impacts the allocation of attention by future researchers and teachers (see Figure 1). In the following sections, we transparently explain the data-gathering process in detail so that other researchers can replicate the analyses in the future.

### 4.1 Articles in Leading Economics Journals

Leading academic journals in economics enjoy authoritative positions. Following prior studies (e.g., Kalaitzidakis et al., 2003; Heckman & Moktan, 2018; Hamermesh 2013, 2018), we focus on the so-called "top5" economics journals: American Economic Review (AER), Quarterly Journal of Economics (QJE), Econometrica, Journal of Political Economy (JPE), and Review of Economic Studies (RES). It is justifiable to say that they form "the core" of the scientific knowledge stock in economics. The majority of the most cited economics papers have been published in these journals.[22] Heckman and Moktan (2018) report that the top5 publications "have a powerful influence on tenure decisions and rates of transition to tenure," and that the "pursuit of T5 publications has become the obsession of the next generation of economists." They also show, using a survey, that the perceptions of young economists are consistent with this view.

Furthermore, several of the most important articles in the field of economic growth have been published in top5 journals. These include Romer (1990), published in JPE; Solow (1956), published in QJE; Grossman and Helpman (1991) and Acemoglu (2002), published in REStud; Aghion & Howitt (1992), published in Econometrica; and Kuznets (1955) and Nelson and Phelps (1966), published in AER. Of these authors Kuznets, Solow, Phelps, and Romer received Nobel Memorial Prizes in Economic Sciences, in 1971, 1987, 2006, and 2018, respectively.

As we focus on analyzing the association between standardization and economic growth, we add the Journal of Economic Growth to complement top5 journals in the sample of economics journals. The Journal of Economic Growth is the leading special journal in the field. Here, the unit of observation is an article in the sample of the mentioned leading economics journals. The search query that we apply in the Scopus database is ALL("standardization" OR "standardisation") AND ALL(technology OR technologies OR technological OR technical) AND ALL("economic growth").

### 4.2 Articles by Leading Researchers of Economic Growth

Leading researchers of economic growth are the key decision makers who make important choices about how the paradigms and trajectories of economic growth research evolve over time (cf. Figure 1). Their allocation of attention directs, in a path-dependent manner, the attention of other researchers.

We analyze whether leading researchers of economic growth have studied the link between standardization and economic growth. We acknowledge that there are various alternative ways to define the leading researchers of economic growth. We limited the sample to editors of Journal of Economic Growth (as of June 2019), which is one of the leading journals in the field, and which has been published since 1996. Its editorial board includes Paul Romer and Paul Krugman, both recipients of the Nobel Memorial Prize in Economics. In addition, most of the editors have written articles to the "Handbook of Economic Growth," which was also edited by Philippe Aghion and Steven Durlauf, editors of the Journal of Economic Growth. There were 34 editors as of June 2019, and Scopus found 2212 documents (incl. articles and books) published by them as of 14 August 2019. The number of documents is biased downward as the Scopus database does not include all older articles (pre-1996). In order to identify documents that focus on standardization and economic growth, we conducted keyword searches in the Scopus database for each author's documents separately using the following search query: ALL("standardization" OR "standardisation") AND ALL(technology OR technologies OR technological OR technical) AND ALL("economic growth"). Each author had on average 65 publications (median: 53) in the Scopus database. Thus, the unit of observation here is an article written by an editor of the Journal of Economic Growth.

### 4.3 Books Related to Economic Growth

Although books are not the major vehicle of scholarly communication in economics (Hamermesh, 2018), they are still often used in teaching as textbooks. Presumably, books on economic growth are an important knowledge source for students of economic growth (cf. Figure 1) since they are often used as coursebooks and enter the syllabi of university courses that focus on economic growth theory.[23] Books may therefore frame the thinking of future economic growth researchers (cf. Kuhn, 1962).

We limit our attention to books published by editors of the Journal of Economic Growth. We inquired whether "standardization" or "standards" occurred in their indexes. The unit of observation is an index of an economic-growth-related book written by an editor of the Journal of Economic Growth.

### 5 FINDINGS

### 5.1 Top Economics Journals

Table 2 reports the numbers of articles published in the leading economics journals that are captured using the specific search terms. The table indicates that there are only a few articles published in top5 journals related to standardization and economic growth. Table 3 lists the articles that are found using the search terms in column 3 of Table 2.

A manual check of the articles in Table 3 reveals that they in fact do not focus on standardization in the sense discussed in this article (i.e., standards development). Articles by Adserà and Ray (1998) and Sákovics and Steiner (2012) are captured by the keyword search because they cite Farrell and Saloner's (1985) article, which has the word "standardization" in its title, instead of actually analyzing the association between standardization and economic growth. Similarly, Acemoglu (2007) cites Farrell and Saloner (1985) but mentions "standardization" also on page 1378 in a footnote: "I assume that the research firm can only choose one technology, which might be, for example, because of the necessity of *standardization* across firms." Also, Jovanovic and Rousseau (2014) does not focus on standards development, although they write on page 864, "Our distinction between extensive investment in new projects and intensive investment in continuing projects is related to one made by Acemoglu, Gancia, and Zilibotti 2012 between innovation and *standardization* costs." Acemoglu and Restrepo (2018) have a labor economics perspective on standardization, as they focus on "standardization of tasks" instead of product standardization and standards development. Finally, Acemoglu et al. (2018) is captured because it refers to Acemoglu et al. (2012), which has the word "standardization" in its title.

To summarize, we did not find one single article published in top5 economics journals that analyses the link between standardization and economic growth. It is worth noting that in the context

Table 2. Leading economics journals

| Journal | Time window | Scopus search queries | | |
|---|---|---|---|---|
| | | ALL("standardization" OR "standardisation") AND ALL(technology OR technologies OR technological OR technical) AND SRCTITLE ("*Journal*") | ALL("standardization" OR "standardisation") AND ALL("economic growth") AND SRCTITLE("*Journal*") | ALL("standardization" OR "standardisation") AND ALL(technology OR technologies OR technological OR technical) AND ALL("economic growth") AND SRCTITLE("*Journal*") |
| AER | 1996-2018 | 12 | 3 | 3 |
| QJE | 1996-2018 | 3 | 0 | 0 |
| JPE | 1996-2018 | 1 | 1 | 1 |
| Econometrica | 1996-2018 | 1 | 1 | 1 |
| REStud | 1996-2018 | 5 | 1 | 0 |
| J Econ Growth | 1996-2018 | 1 | 1 | 1 |

Notes: Time window is limited to 1996-2018.

Table 3. Articles related to standardization and economic growth

| | Article | Author(s) | Year | Journal Issue | Scopus citations |
|---|---|---|---|---|---|
| 1 | History and coordination failure | Adserà, A. & Ray, D. | 1998 | Journal of Economic Growth 3(3), 267-276 | 26 |
| 2 | Equilibrium bias of technology | Acemoglu, D. | 2007 | Econometrica 75(5), 1371-1409 | 80 |
| 3 | Who matters in coordination problems? | Sákovics, J. & Steiner, J. | 2012 | American Economic Review 102(7), 3439-3461 | 22 |
| 4 | Extensive and intensive investment over the business cycle | Jovanovic, B. & Rousseau, P.L. | 2014 | Journal of Political Economy 122(4), 863-908 | 7 |
| 5 | The race between man and machine: Implications of technology for growth, factor shares, and employment | Acemoglu, D. & Restrepo, P. | 2018 | American Economic Review 108(6), 1488-1542 | 16 |
| 6 | Innovation, reallocation, and growth | Acemoglu, D., Akcigit, U., Alp, H., Bloom, N. & Kerr, W. | 2018 | American Economic Review 108(11), 3450-3491 | 13 |

Notes: Based on Scopus results. As of 14 Aug 2019.

of labor economics, the concept of standardization is more related to "standardization of tasks" (e.g., Acemoglu et al., 2012, 2018) than product or system standardization or technical specifications.

## 5.2 Leading Researchers of Economic Growth

As of August 2019, the Journal of Economic Growth had published no articles that (1) focus on technology standardization (see Section 4.1), (2) list "standard," "standards," or "standardization" as keywords, or (3) list "L15" as a JEL classification code.[24] Using economics jargon: It seems that the existing research on standardization has not been used as an input in the production of economic growth theory. Therefore, we focus in this and the next section on more deeply analyzing the attention allocated to standardization by the editors of the Journal of Economic Growth.

Table 4 reports the number of articles by leading economic growth researchers[25] that mention specific keywords related to standardization and economic growth. It is notable that only six out of 34 editors (~18%) are affiliated with European universities, while 26 are affiliated with American universities, and three with an Israeli university.

Table 5 lists the articles that are authored by editors of the Journal of Economic Growth and are captured by the specific search terms related to standardization and economic growth presented in Section 4.2. Note that there is a significant overlap between articles in Table 3 and Table 5, as four articles can be found in both tables. Notably, Acemoglu is an author in seven of the 11 identified articles. We explained already above why Adserà and Ray (1998), Acemoglu (2007), Acemoglu (2018), and Acemoglu and Restrepo (2018) are captured. A closer look at the contents of the seven other articles reveals why the search query captures them. Durlauf (2005) is captured because its reference list includes Farrell and Saloner (1985), and Acemoglu et al. (2015) and Zilibotti (2017) are captured because they cite Acemoglu et al. (2012). Acemoglu and Akcigit (2012) is captured as it cites Acemoglu et al. (2012) in footnote 7, and Acemoglu (2012) is an introduction to the special issue of the Journal of Economic Theory which includes and introduces Acemoglu et al. (2012). Acemoglu et al. (2012) focuses on standardization but from a slightly differing theoretical "labor economics" perspective. In other words, Aghion et al. (2009) is the only article by an editor of the Journal of Economic Growth that actually discusses standardization from the perspective of this article. Interestingly, their perspective on standardization seems to be relatively negative, as they write (p.689): "In network industries, and in product markets characterized by network externality effects, a policy stance of avoiding deliberate standard-setting is not a strategy sufficient to prevent regrettable *standardization* outcomes, in which industries are 'locked in' to an inferior technical system that proves costly to abandon." They also highlight on page 689 that "Perhaps the most productive question to ask is how we can identify situations in which, at some future time, most technology users would look back and agree that they would have been better off had they converged on the adoption of an alternative technical option (David, 1987)."

## 5.3 Books Related to Economic Growth

Table 6 shows that most economic growth theory textbooks do not mention "standardization" in their indexes. Most prominently, the authoritative "Handbook of Economic Growth" (Aghion & Durlauf, 2006, 2014) is among that majority. Yet, a few exceptions are found. David Weil (2012) mentions standards briefly but focuses on government-imposed standards on page 305:

*"Excessive standards – Governments impose standards on all sorts of goods that are sold in their countries, ranging from regulations designed to protect public health (e.g., pure-food standards) to requirements that enable different pieces of equipment to work together. Often, however, standards are used to keep foreign products out of the market. For example, Israel, with a population of only 6 million, requires the use of an electrical plug that is unique in the world, to give an advantage to local manufacturers of electrical equipment."*

## Table 4. Leading researchers of economic growth

| Editor | Affiliations | Number of documents in Scopus** | Number of articles published in Top5** | Share of Top5 articles of all documents in Scopus** | Coauthors** | Number of citing documents | Articles focusing on standardization and economic growth*** |
|---|---|---|---|---|---|---|---|
| Oded Galor* | Brown University; Hebrew University | 52 | 17 | 32.69% | 21 | 4644 | 0 |
| Daron Acemoglu | MIT | 238 | 74 | 31.09% | 126 | 20408 | 7 |
| Philippe Aghion | Harvard University | 148 | 27 | 18.24% | 146 | 10821 | 1 |
| Ufuk Akcigit | University of Chicago | 19 | 11 | 57.89% | 20 | 295 | 2 |
| Alberto Alesina | Harvard University | 121 | 28 | 23.14% | 86 | 15857 | 0 |
| Quamrul Ashraf | Williams College | 12 | 3 | 25.00% | 8 | 457 | 0 |
| Roland Benabou | Princeton University | 15 | 10 | 66.67% | 7 | 3442 | 0 |
| Jess Benhabib | New York University | 89 | 9 | 10.11% | 44 | 4194 | 0 |
| Jagdish Bhagwati | Columbia University | 140 | 13 | 9.29% | 48 | 3780 | 0 |
| Francesco Caselli | Harvard University | 29 | 10 | 34.48% | 20 | 2810 | 0 |
| Carl-Johan Dalgaard | University of Copenhagen | 29 | 1 | 3.45% | 16 | 817 | 0 |
| Matthias Doepke | UCLA | 26 | 11 | 42.31% | 12 | 976 | 0 |
| Steven Durlauf | University of Wisconsin | 102 | 11 | 10.78% | 64 | 5389 | 2 |
| William Easterly | New York University | 99 | 7 | 7.07% | 47 | 10295 | 0 |
| James Fenske | University of Warwick | 22 | 2 | 9.09% | 11 | 232 | 0 |
| Gene Grossman | Princeton University | 89 | 32 | 35.96% | 23 | 8719 | 0 |
| Vernon Henderson | Brown University | 99 | 15 | 15.15% | 56 | 5730 | 0 |
| Peter Howitt | Brown University | 66 | 14 | 21.21% | 41 | 3768 | 0 |
| Charles Jones | Stanford University | 31 | 16 | 51.61% | 8 | 6332 | 0 |
| Paul Krugman | Princeton University | 106 | 9 | 8.49% | 27 | 12972 | 0 |
| Ross Levine | Brown University | 84 | 6 | 7.14% | 39 | 16628 | 0 |
| Stelios Michalopoulos | Brown University | 15 | 5 | 33.33% | 9 | 517 | 0 |
| Omer Moav | Hebrew University | 21 | 6 | 28.57% | 13 | 1311 | 0 |
| Joel Mokyr | Northwestern University | 84 | 3 | 3.57% | 40 | 2977 | 1 |
| Torsten Persson | IIES, Stockholm University | 77 | 22 | 28.57% | 33 | 5352 | 0 |
| Debraj Ray | New York University | 96 | 27 | 28.13% | 107 | 3084 | 1 |
| Paul Romer | Stanford University | 25 | 9 | 36.00% | 10 | 8993 | 0 |
| Nancy Stokey | University of Chicago | 29 | 5 | 17.24% | 16 | 2938 | 0 |
| Jonathan Temple | Bristol University | 39 | 1 | 2.56% | 18 | 2518 | 0 |
| Hans-Joachim Voth | Pompeu Fabra University | 54 | 12 | 22.22% | 21 | 1020 | 0 |
| Romain Wacziarg | UCLA | 25 | 5 | 20.00% | 16 | 4322 | 0 |
| David Weil | Brown University | 62 | 17 | 27.42% | 40 | 9334 | 0 |
| Joseph Zeira | Hebrew University | 19 | 4 | 21.05% | 12 | 1689 | 0 |
| Fabrizio Zilibotti | Stockholm University | 50 | 12 | 24.00% | 32 | 2877 | 4 |

Notes: *Editor in chief ** As of 14th August 2019 ***Search in Scopus: Among the documents published by the selected author, ALL("standardization" OR "standardisation") AND ALL(technology OR technologies OR technological OR technical) AND ALL("economic growth")

**Table 5. Articles by editors of the Journal of Economic Growth**

| Article | Authors* | Year | Journal | Citations in Scopus** |
|---|---|---|---|---|
| History and coordination failure | Adserà, A., **Ray, D.** | 1998 | Journal of Economic Growth | 26 |
| Complexity and empirical economics | **Durlauf, S.N.** | 2005 | Economic Journal 115(504), pp. F225-F243 | 80 |
| Equilibrium bias of technology | **Acemoglu, D.** | 2007 | Econometrica 75(5), pp. 1371-1409 | 80 |
| Science, technology and innovation for economic growth: Linking policy research and practice in 'STIG Systems' | **Aghion, P.**, David, P., Foray, D. | 2009 | Research Policy 38(4), 681-693 | 105 |
| Intellectual property rights policy, competition and innovation | **Acemoglu, D., Akcigit, U.** | 2012 | Journal of the European Economic Association 10(1), pp. 1-42 | 70 |
| Introduction to economic growth | **Acemoglu, D.** | 2012 | Journal of Economic Theory 147(2), pp. 545-550 | 7 |
| Competing engines of growth: Innovation and standardization | **Acemoglu, D.**, Gancia, G., **Zilibotti, F.** | 2012 | Journal of Economic Theory 147(2), pp. 570-601.e3 | 48 |
| Offshoring and directed technical change | **Acemoglu, D.**, Gancia, G., **Zilibotti, F.** | 2015 | American Economic Journal: Macroeconomics 7(3), pp. 84-122 | 31 |
| Growing and slowing down like China | **Zilibotti, F.** | 2017 | Journal of the European Economic Association 15(5), pp. 943-988 | 2 |
| Innovation, reallocation, and growth | **Acemoglu, D.**, **Akcigit, U.**, Alp, H., Bloom, N., Kerr, W. | 2018 | American Economic Review 108(11), pp. 3450-3491 | 13 |
| The race between man and machine: Implications of technology for growth, factor shares, and employment | **Acemoglu, D.**, Restrepo, P. | 2018 | American Economic Review 108(6), pp. 1488-1542 | 16 |

Notes: *Editors of J of Econ Growth bolded. **As of 14 Aug 2019

Mokyr (2002) mentions "standardization" a few times, for instance, on page 111:

*"Modularization was closely related to standardization, making all products of particular type conform to a uniform standard. Standardization, much like modularization, helped not just during the production stage of output but also in the maintenance of durable equipment. Whoever could repair one Model T could repair any Model T."*

These observations suggest that standardization plays no focal role in economic growth books. This indicates that researchers of economic growth do not generally consider standardization to be an important factor affecting technological progress and economic growth.

A comprehensive keyword search within the content of these books is left for future research. However, we conducted a keyword search using the search term "standardization" among the articles that are included in volumes 1 and 2 of the Handbook of Economic Growth (Aghion & Durlauf, 2006, 2014). The "Handbook of Economic Growth" which is edited by world-leading researchers of economic growth, could be considered to represent the stage of the current scientific paradigm. We found one article that included the word "standardization": Ventura (2005) mentions "Advances in telecommunications technology and the standardization of software allow producers around the world to combine physical and human capital located in different regions in a single production process," but it does not discuss standardization more extensively. This observation further corroborates the currently missing link between standardization and economic growth theory.

To summarize, these findings suggest that there is a research gap regarding the association between standardization and economic growth. The current paradigm of economic growth theory does not incorporate standardization. Top economics journals published no articles related to the topic between 1996 and 2018. The leading journal in its field, the Journal of Economic Growth, has not yet touched upon standardization, and most researchers on its editorial board have allocated only little attention to standardization. Finally, the concept of standardization is not well-specified and, recently, seems to be more often related to "standardization of tasks."

## 5.4 Limitations and Avenues for Future Research

The current analysis has several limitations. For instance, it includes only a small portion of leading economics journals and excludes several journals that have published important articles related to standardization (e.g., Journal of Economics and Management Strategy, The Economic Journal, European Journal of Political Economy). Second, the applied keyword search methodology is not necessarily the most accurate one, and additional robustness checks could be conducted. Finally, the current analysis is just a snapshot of a specific research gap at a specific point in time and thus it becomes obsolete quickly as the literature on standardization continue to grow.

Initially, we planned to use the Journal of Economic Literature (JEL) classifications codes in identifying standardization-related articles (cf. Cherrier, 2017). JEL classification code L15 is "Information and Product Quality: Standardization and Compatibility" and, according to JEL guidelines it "includes studies on standardization and on compatibility, which reduces the problems associated with the information-product-quality nexus."[27] The JEL classification also lists "compatibility," "standardization," and "ISO" as keywords belonging to L15. However, the use of JEL classification codes is not always consistent. As an example, whereas Baron and Schmidt (2017) is classified into categories E32, E22, O33, O47, and L15, Blind and Jungmittag (2008) is classified into O41, O52, O11, and E13. There is no overlap despite the fact that both articles analyze the macroeconomic impact of standards. When looking at keywords, there is a similar lack of overlap, but both articles have "standards" or "standardization" as keywords.[28] Due to these inconsistencies, this alternative JEL-classification search option is left for future studies and possible replications and updates.

As illustrated in Section 3.2, accumulated economics research and textbooks affect the curricula of economic growth courses. Future studies could extend the analysis to reviewing the role of standardization in economics courses. Acemoglu (2013) has recommended that economics instructors should spend more time on teaching economic growth at the undergraduate level and also on emphasizing the importance of technology as the key determinant of economic growth. The same growth course could be accompanied by a brief review of standardization so that economics students can acknowledge the importance of standards. There already exist multiple initiatives to increase awareness of the importance of standards (de Vries & Egyedi, 2007; Choi & de Vries, 2011; Blind & Dreschler, 2017).

Economics has shifted over time, increasingly from theoretical modelling to empirical analysis (Hamermesh, 2013). As there are ever-increasing data on standardization available for researchers (e.g., Baron & Spulber, 2018; Baron & Gupta, 2018), it is expected that there will also be more publications on the topic and empirical economists will begin to allocate more attention to standards. Future research could focus on analyzing and quantifying the macro-level economic impact of standards. Evidence-based policies require rigorous empirical analysis and, presumably, the welfare effects of standardization will receive more attention in the future.

**Table 6. Books related to economic growth**

| Authors | Book | Publisher | Year | Index includes "standards" or "standardization" |
|---------|------|-----------|------|--------------------------------------------------|
| **Joel Mokyr** | The Lever of Riches: Technological Creativity and Economic Progress | Oxford University Press | 1990 | - |
| **Gene Grossman** and Elhanan Helpman | Innovation and Growth in the Global Economy | MIT Press | 1991 | - |
| **Gene Grossman** | Economic Growth: Theory and Evidence | Edward Elgar Publishing | 1996 | No access |
| **Philippe Aghion** and **Peter Howitt** | Endogenous Growth Theory | MIT Press | 1997 | - |
| **Philippe Aghion** and Jeffrey Williamson | Growth, Inequality, and Globalization: Theory, History, and Policy | Cambridge University Press | 1999 | - |
| **William Easterly** | The Elusive Quest for Growth: Economists' Adventures and Misadventures in the Tropics (The MIT Press) | MIT Press | 2002 | - |
| **Philippe Aghion** and Rachel Griffith | Competition And Growth: Reconciling Theory And Evidence | MIT Press | 2005 | - |
| **Philippe Aghion** and Abhijit Banerjee | Volatility and Growth | Oxford University Press | 2005 | - |
| Asli Demirgüç-Kunt and **Ross Levine** | Financial Structure and Economic Growth | MIT Press | 2005 | - |
| **Philippe Aghion** and **Steven Durlauf** (eds.) | Handbook of Economic Growth 1A | Elsevier/North Holland | 2006 | - |
| **Philippe Aghion** and **Steven Durlauf** (eds.) | Handbook of Economic Growth 1B | Elsevier/North Holland | 2006 | - |
| **Daron Acemoglu** | Introduction to Modern Economic Growth | Princeton University Press | 2008 | - |
| **Philippe Aghion** and **Peter Howitt** | Economics of growth | MIT Press | 2008 | - |
| **Oded Galor** | Unified Growth Theory | Princeton University Press | 2011 | - |
| **Joel Mokyr** | The Gifts of Athena: Historical Origins of the Knowledge Economy | Princeton University Press | 2011 | standardization, 58, 60, 63, 111, 229, 257 |
| **David Weil** | Economic growth (3rd ed.) | Routledge | 2012 | Standards, as trade barriers, 305 |
| **Charles I. Jones** and Dietrich Vollrath | Introduction to Economic Growth (3rd ed.) | W. W. Norton & Company | 2013 | - |
| **Jagdish Bhagwati** and Arvind Panagariya | Why Growth Matters: How Economic Growth in India Reduced Poverty and the Lessons for Other Developing Countries | PublicAffairs | 2014 | |
| **Philippe Aghion** and **Steven Durlauf** (eds.) | Handbook of Economic Growth 2A | Elsevier/North Holland | 2014 | - |
| **Philippe Aghion** and **Steven Durlauf** (eds.) | Handbook of Economic Growth 2B | Elsevier/North Holland | 2014 | - |
| **Francesco Caselli** | Technology Differences over Space and Time | Princeton University Press | 2016 | - |
| **Joel Mokyr** | A Culture of Growth: The Origins of the Modern Economy | Princeton University Press | 2016 | - |
| Asli Demirgüç-Kunt and **Ross Levine** | Finance and Growth | International Library of Critical Writings in Economics | N/A | No access |

Notes: Economic-growth-related books published by the editors of the Journal of Economic Growth as of August 2019.

## 6 CONCLUDING REMARKS

The main findings of this article are the following. No article has analyzed the link between standardization and economic growth in top5 economics journals and the Journal of Economic Growth. A representative sample of leading researchers of economic growth has allocated only negligible attention to the link between standardization and economic growth. Economic growth theory textbooks and closely related books only occasionally mention standardization. Based on these findings, it is plausible to conclude that the current paradigm of economic growth theory neglects standardization. We confirm the observation that there are very few academic studies in the field of economics that analyze the contribution of standardization to economic growth (Blind et al., 2005; Blind & Jungmittag, 2008; Baron & Schmidt, 2017).

Existing empirical evidence suggests that standards may have significant economic impacts. Yet, the role of standardization as a factor in economic growth and development has, thus far, been neglected. These observations indicate that mainstream economic growth researchers have not considered standardization to be an important determinant of economic growth and prosperity. This lack of attention may have significant implications. Since university teaching is research-based, with the limited accumulated research on the role of standards in technological change there exists a risk that standardization will receive little attention in the future as well. Standards matter for technological progress, productivity, and economic growth. Economic growth researchers could further shed light on the black box of technological progress by allocating more attention to standardization.

## ACKNOWLEDGEMENTS

Earlier versions of this paper have been presented at the XXXVI Summer Seminar of Finnish Economists, JSBE Research Seminar, ETLA brownbag seminar, the Finnish Economic Association's 42nd Annual Meeting and it was accepted for presentation at the European Academy for Standardisation (EURAS) conference in Glasgow (postponed due to COVID-19). We thank Samira Ranaei, Johan Willner, Peter Swann and several anonymous reviewers for helpful comments. Joakim Wikström provided excellent research assistance. The views expressed herein are those of the authors and do not necessarily reflect the views of their employers. All errors are our own.

## ENDNOTES

[1]   https://www.iso.org/members.html Last accessed on 1 August 2019.

[2]   Although some national standards organizations are also very important.

[3]   One way to describe the difference is to note that standard setting and development constitute a competition between standards and within standards, respectively.

[4]   Technological progress and technological change are used interchangeably throughout the paper. We do not use the term "technical change."

[5]   The British Standards Institute, BSI, was the world's first National Standards Body, formed in 1901. The German Institute for Standardization, DIN, was founded in 1917, and the American National Standards Institute, ANSI, was founded in 1918.

[6]   Akcigit (2017) provides an overview of the past, present, and future of economic growth theory, and Jones (2016) an overview of the facts. Chu (2018) provides an overview of economic growth teaching curricula. A glimpse into the most recent version of the "Handbook of Economic Growth" provides an overview of the growth factors to which growth researchers have recently allocated their attention (Aghion et al., 2014).

[7]   David and Greenstein (1990) provides an overview.

[8]   See also Veblen (1904) for early contributions on the "economics of standardization" field.

[9]   Standardized written and spoken languages can also be considered as the ultimate "GPTs." Mokyr (2002, p. 58) notes: "For communication between individuals to occur, a common terminology is essential. Language is the ultimate general purpose technology, to use Bresnahan and Trajtenberg (1993) well-known term. It provides the technology that creates others." As a more recent example of language standardization: Emojis, special pictorial symbols, are standardized by Unicode Emoji Subcommittee.

[10]  "If one views the Internet as a general purpose technology, these standard-setting organizations may provide a forum where GPT-producers can interact with application-sector innovators in an effort to internalize the vertical (from GPT to application) and horizontal (among applications) externalities implied by complementarities in innovation across sectors, as modeled in Bresnahan and Trajtenberg (1995)." (Simcoe 2015, p.26).

[11]  Dosi (1982) defines "technological paradigm" as procedures, definitions of the "'relevant' problems, and as the specific knowledge related to their solution, and "technological trajectory" as the direction of advance within a technological paradigm.

[12]  For instance, regarding the standardization of the Internet, Simcoe notes that "Consensus standardization within SSOs (specifically IETF and W3C, as described below) is arguably the dominant mode of coordinating the design decisions and the supply of new interfaces on the modern Internet" (2015 p.26).

[13]  https://www.econlib.org/library/Enc/EconomicGrowth.html Last accessed 16 Aug 2019.

[14]  Quantification is challenging, and existing country comparisons rely on a "stock of standards" proxy variable when analyzing the association between standards and economic growth (e.g., Blind & Jungmittag, 2008).

[15]  Rysman and Simcoe (2008): "The large difference in baseline citation rates suggests that SSOs perform well in selecting important technologies. If we are willing to place a causal interpretation on the disclosure effect, these results also imply that SSOs increase the significance of standardized technology through formal endorsement and other efforts to promote industry coordination."

[16]  According to Leonard (2002): "Scientific rules, and the means for their enforcement, constitute the invisible-hand mechanism, so that scientific rules (sometimes) induce interested scientific actors with worldly goals to make epistemically good choices."

[17]  Division of labor in economics has been studied from various perspectives in prior literature (de Langhe, 2010).

[18]  For herding behavior of researchers, see, e.g., Volume 20 Issue 1 of the Journal of Economic Methodology.

[19]  Kitcher (1990) uses the concept (cognitive division of labor).

[20]  Interestingly, there exists empirical evidence that preferences and self-selection seem to be gender-specific, with the shares of women and men differing across economic research topics (Chari & Goldsmith-Pinkham, 2018).

[21]  These research articles are sometimes called "sleeping beauties" (van Raan, 2004).

[22]  See https://ideas.repec.org/top/top.item.nbcites.html Accessed on 29 February 2020.

[23]  E.g., "Economic growth" course by prof. Daron Acemoglu at MIT in 2016, https://ocw.mit.edu/courses/economics/14-452-economic-growth-fall-2016/syllabus/ uses Acemoglu (2016) as a textbook.

[24]  L15 "Information and Product Quality: Standardization and Compatibility" https://www.aeaweb.org/jel/guide/jel.php. The only keyword that was found to contain the term "standards" was "International Labor Standards."

[25]     Interestingly, economic growth research seems to be male-dominated, as there is only one woman on the editorial board. This observation is consistent with findings of Chari and Goldsmith-Pinkham (2018).

[26]     In addition to these articles, the search query captures three book chapters that are excluded from this table.

[27]     See https://www.aeaweb.org/jel/guide/jel.php Accessed on 27 July 2019

[28]     The keywords of Baron and Schmidt (2017) are "technology adoption; business cycle dynamics; standardization; aggregate productivity; Bayesian vector autoregressions." Blind and Jungmittag (2008) listed as keywords "Growth; Innovation; Standards; Patents; Country effects; Industry effects."